\documentclass[a4paper,11pt]{article}
\usepackage{pifont}
\date{}

\usepackage{makeidx}
\usepackage[centertags]{amsmath}
\usepackage{amsfonts}
\usepackage{amssymb}
\usepackage{amsthm}
\usepackage[dvips]{epsfig}
\usepackage{color}
\usepackage{subfigure}

\makeatletter \@addtoreset{equation}{section} \makeatother

\begin{document}
\title{\bf Periodic Solutions for Circular Restricted 4-body
Problems with Newtonian Potentials \footnote{Supported by National
Natural Science Foundation of China.}}

\author{{ Xiaoxiao Zhao,\ and \ Shiqing Zhang}\\
{\small Yangtze Center of Mathematics and College of Mathematics, Sichuan University,}\\
{\small Chengdu 610064, People's Republic of China}} \maketitle
\begin{quote}

{\bf Abstract:} We study the existence of non-collision periodic
solutions with Newtonian potentials for the following planar
restricted 4-body problems: Assume that the given positive masses
$m_{1},m_{2},m_{3}$ in a Lagrange configuration move in circular
obits around their center of masses, the sufficiently small mass
moves around some body. Using variational minimizing methods, we
prove the existence of minimizers for the Lagrangian action on
anti-T/2 symmetric loop spaces. Moreover, we prove the minimizers
are non-collision periodic solutions with some fixed wingding
numbers.


2000 AMS Subject Classification 34C15, 34C25, 58F.
\end{quote}

\section{Introduction and Main Results}
\ \ \ \ \ \ In this paper, we study the planar circular restricted
4-body problems with Newtonian potentials. Suppose points of
positive masses $m_{1},m_{2},m_{3}$ move in a plane of their
circular orbits $q_{1}(t),q_{2}(t), q_{3}(t)$ and the center of
masses is at the origin; suppose the sufficiently small mass point
does not influence the motion of $m_{1},m_{2},m_{3}$, and  moves in
the plane for the given masses $m_{1},m_{2},m_{3}$.

It is well-known that $q_{1}(t),q_{2}(t), q_{3}(t)$ satisfy the
Newtonian equations:
\begin{equation}\label{e1}
m_{i}\ddot{q_{i}}=\frac{\partial U}{\partial q_{i}},\ \ \ \ i=1,2,3,
\end{equation}
where
\begin{equation}\label{e2}
U=\sum\limits_{1\leq i< j\leq 3}\frac{m_{i}m_{j} }{| q_{i}-q_{j}|}.
\end{equation}
Without loss of generality, we assume that there exists
$\theta_{1},\theta_{2},\theta_{3}\in[0,2\pi)$ such that the planar
circular orbits are
\begin{equation}\label{e3}
q_{1}(t)=r_{1}e^{\sqrt{-1}\frac{2\pi}{T} t}e^{\sqrt{-1}\theta_{1}},\
\ q_{2}(t)=r_{2}e^{\sqrt{-1}\frac{2\pi}{T}
t}e^{\sqrt{-1}\theta_{2}}, \ \
q_{3}(t)=r_{3}e^{\sqrt{-1}\frac{2\pi}{T} t}e^{\sqrt{-1}\theta_{3}},
\end{equation}
where the radius $r_{1}, r_{2},r_{3}$ are positive constants
depending on $m_{i}(i=1,2,3)$ and $T$ (see Lemma 2.6). We also
assume that
\begin{equation}\label{e4}
m_{1}q_{1}(t)+m_{2} q_{2}(t)+m_{3} q_{3}(t)=0
\end{equation}
and
\begin{equation}\label{e5}
|q_{i}-q_{j}|=l, \ \ 1\leq i\neq j\leq3,
\end{equation}
where the constant $l>0$ depends on $m_{i}(i=1,2,3)$ and $T$ (see
Lemma 2.5).

The orbit $q(t)\in R^{2}$ for sufficiently small mass is governed by
the gravitational forces of $m_{1},m_{2},m_{3}$ and therefore it
satisfies the following equation
\begin{equation}\label{e6}
\ddot{q}=\sum\limits_{i=1}^{3}\frac{m_{i}(q_{i}-q) }{|
q_{i}-q|^{3}}.
\end{equation}
\ \ \ \ \ \ \ For $N$-body problems, there are many papers concerned
with the periodic solutions by using variational methods, see
[1-9,13-16,18] and the references therein. In \cite{r3},
Chenciner-Montgomery proved the existence of the remarkable
figure-``8'' type periodic solution for planar Newtonian 3-body
problems with equal masses. Marchal\cite{r6} studied the fixed end
problem for Newtonian n-body problems and proved the minimizer for
the Lagrangian action has no interior collision. Especially, in
\cite{r8}, Sim\'{o} used computer to discover many new periodic
solutions for Newtonian n-body problems. Zhang-Zhou[13-15]
decomposed the Lagrangian action for n-body problems into some sum
for two body problems and [14,15] avoid collisions by comparing the
lower bound for the Lagrangian action on the symmetry collision
orbits and the upper bound for the Lagrangian action on test orbits
in some cases.

Motivated by the above works, we use variational methods to study
the circular restricted 3+1-body problem with some fixed wingding
numbers and some masses.

For the readers' conveniences, we recall the definition of the
winding number, which can be found in many books on the
classical differential geometry.

\vspace{0.4cm}\textbf{Definition 1.1} \  Let $\Gamma: x(t), t \in
[a,b]$ be an given oriented continuous closed curve, and $p$ be a
point of the plane not on the curve. Then, the mapping $\varphi:
\Gamma\rightarrow S^{1}$, given by
\begin{equation*}
\varphi(x(t)) = \frac{x(t)-p}{|x(t)-p|},\ \ \ t \in [a,b]
\end{equation*}
is defined to be the position mapping of the curve $\Gamma$ relative
to $p$, when the point on $\Gamma$ goes around the curve once, its
image point $\varphi(x(t))$ will go around $S^{1}$ a number of
times, this number is called the winding number of the curve
$\Gamma$ relative to $p$, and we denote it by $deg(\Gamma,p)$. If
$p$ is the origin, we write $deg\Gamma$.

Define
\begin{equation*}
W^{1,2}(R/TZ,R^{2})=\bigg\{x(t)\Big| x(t),\dot{x}(t)\in
L^{2}(R,R^{2}), \ x(t+T)=x(t) \bigg\}.
\end{equation*}
The norm of $W^{1,2}(R/TZ,R^{2})$ is
\begin{equation}\label{e7}
\|x\|=\Big[\int_{0}^{T}|x|^{2}dt\Big]^{\frac{1}{2}}+\Big[\int_{0}^{T}
|\dot{x}|^{2}dt\Big]^{\frac{1}{2}}.
\end{equation}
The functional corresponding to the equation (\ref{e6}) is
\begin{equation}\label{e8}
f(q)=\int_{0}^{T}\Big[\frac{1}{2}|\dot{q}|^{2}+\sum\limits_{i=1}^{3}\frac{m_{i}
}{| q-q_{i}|}\Big]dt,\ \ \ \ q\in \Lambda_{\pm},
\end{equation}
where
\begin{equation*}
\Lambda_{-}=\left\{ q\in W^{1,2}(R/TZ,R^{2})\bigg|
\begin{array}{c}
q(t+\frac{T}{2})=-q(t),\
deg(q-q_{1})=-1,\\
q(t)\neq q_{i}(t),\  \forall t\in[0,T],i=1,2,3\ \ \ \ \ \ \
\end{array}
\right\}.
\end{equation*}
and
\begin{equation*}
\Lambda_{+}=\left\{ q\in W^{1,2}(R/TZ,R^{2})\bigg|
\begin{array}{c}
q(t+\frac{T}{2})=-q(t),\
deg(q-q_{1})=1,\\
q(t)\neq q_{i}(t),\  \forall t\in[0,T],i=1,2,3\ \ \ \ \ \ \
\end{array}
\right\}.
\end{equation*}

Our main results are the following:

\vspace{0.4cm}\textbf{Theorem 1.1} \ Let $T=1$, for the values of
$m_{1},m_{2},m_{3}$ given in Table 1 with $M=1$, the minimizer of
$f(q)$ on the closure $\overline{\Lambda}_{-}$ of $\Lambda_{-}$ is a
non-collision 1-periodic solution of (\ref{e6}); for the values of
$m_{1},m_{2},m_{3}$ given in Table 2 with $m_{1}=m_{2}=m_{3}=1$, the
minimizer of $f(q)$ on $\overline{\Lambda}_{-}$ is a non-collision
1-periodic solution of (\ref{e6}).

\vspace{0.4cm}{\textbf{Remark 1}}\ \ In proving Theorem 1, we need
to use test functions. We find that if the test functions are
circular orbits, we can not get the desired results on
$\overline{\Lambda}_{-}$. Therefore, we select elliptic orbits as
test functions.

\vspace{0.4cm}\textbf{Theorem 1.2} \ Let $T=1$, for the values of
$m_{1},m_{2},m_{3}$ given in Table 3 with $M=1$, the minimizer of
$f(q)$ on the closure $\overline{\Lambda}_{+}$ of $\Lambda_{+}$ is a
non-collision 1-periodic solution of (\ref{e6}); for the values of
$m_{1},m_{2},m_{3}$ given in Table 4 with $m_{1}=m_{2}=m_{3}=1$, the
minimizer of $f(q)$ on $\overline{\Lambda}_{+}$ is a non-collision
1-periodic solution of (\ref{e6}).

\vspace{0.4cm}{\textbf{Remark 2}}\ \ When we take elliptic orbits as
test functions, we find that the biggest symmetric space is the
anti-T/2 symmetric loop space if the wingding number $n$ is
odd($n=\pm1,\pm3,\cdots$); we can not find suitable symmetric space
if the wingding number is even. When the wingding number $n\neq\pm1$
and we take circular orbits as test functions, we find that the
biggest symmetric space is
\begin{equation*} \Lambda=\left\{
q\in W^{1,2}(R/TZ,R^{2})\bigg|
\begin{array}{c}
q(t+\frac{T}{|n-1|})=R(|n-1|)q(t),\
deg(q-q_{1})=n,\\
q(t)\neq q_{i}(t),\  \forall t\in[0,T],i=1,2,3\ \ \ \ \ \ \
\end{array}
\right\},
\end{equation*}
where
\begin{equation*}
R(|n-1|)=\left(\begin{array}{ccc}
cos\frac{2\pi}{|n-1|} & -sin\frac{2\pi}{|n-1|}\\
sin\frac{2\pi}{|n-1|} &  cos\frac{2\pi}{|n-1|}
\end{array}
\right)\in SO(2)
\end{equation*}
is a counter-clockwise rotation of angle $\frac{2\pi}{|n-1|}$ in
$R^{2}$. But the Lagrangian actions on the circular test orbits are
bigger than the lower bound for the Lagrangian actions on collision
symmetric orbits. Hence we consider the anti-T/2 symmetric loop
spaces $\Lambda_{\pm}$.

\section{Preliminaries}
In this section,  we will list some basic Lemmas and inequality for
proving our Theorems 1.1 and 1.2.

\vspace{0.4cm}{\textbf{Lemma 2.1}}(Tonelli\cite{r1},\cite{r11}) Let
$X$ be a reflexive Banach space, $S$ be a weakly closed subset of
$X$, $f:S\rightarrow R\cup \{+\infty\}$. If $f\not\equiv +\infty$ is
weakly lower semi-continuous and coercive($f(x)\rightarrow +\infty$
as $\|x\|\rightarrow +\infty$), then $f$ attains its infimum on $S$.

\vspace{0.4cm}\textbf{Lemma 2.2}(Poincare-Wirtinger
Inequality\cite{r10})\ \ Let $q\in W^{1,2}(R/TZ,R^{K})$ and
$\int_{0}^{T}q(t)dt=0$, then
\begin{equation*}
\int_{0}^{T}|q(t)|^{2}dt\leq\frac{T^{2}}{4\pi^{2}}\int_{0}^{T}|\dot{q}(t)|^{2}dt.
\end{equation*}

\vspace{0.4cm}{\textbf{Lemma 2.3}}(Palais's Symmetry
Principle(\cite{r12}))\ Let $\sigma$ be an orthogonal representation
of a finite or compact group $G$, $H$ be a real Hilbert space,
$f:H\rightarrow R$ satisfies $f(\sigma\cdot x)=f(x),\forall\sigma\in
G,\forall x\in H$.

Set $F=\{x\in H|\sigma\cdot x=x,\ \forall \sigma\in G\}$. Then the
critical point of $f$ in $F$ is also a critical point of $f$ in $H$.

\vspace{0.4cm}{\textbf{Remark 2.1}}\ \ By Palais's Symmetry
Principle and the perturbation invariance for wingding numbers, we
know that the critical point of $f(q)$ in $\Lambda_{\pm}$ is a
periodic solution of Newtonian equation (\ref{e6}).

\vspace{0.4cm}{\textbf{Lemma 2.4}}

(1)(Gordon's Theorem\cite{r17}) Let $x\in W^{1,2}([t_{1},
t_{2}],R^{K})$ and $x(t_{1}) = x(t_{2}) = 0$. Then for any $a
> 0$, we have
\begin{equation*}
\int_{t_{1}}^{t_{2}}(\frac{1}{2}|\dot{x}|^{2}+\frac{a}{|x|})dt
\geq\frac{3}{2}(2\pi)^{2/3}a^{2/3}(t_{2}-t_{1})^{1/3}.
\end{equation*}

(2)(Long-Zhang\cite{r18}) Let $x\in W^{1,2}(R/TZ,R^{K}),\int_{0}^{T}
xdt = 0$, then for any $a > 0$, we have
\begin{equation*}
\int_{0}^{T}(\frac{1}{2}|\dot{x}|^{2}+\frac{a}{|x|})dt
\geq\frac{3}{2}(2\pi)^{2/3}a^{2/3}T^{1/3}.
\end{equation*}

\vspace{0.4cm}{\textbf{Lemma 2.5}}\ \ Let $M=m_{1}+m_{2}+m_{3}$, we
have $l=\sqrt[3]{\frac{MT^{2}}{4\pi^{2}}}$.

\vspace{0.4cm}{\textbf{Proof.}}\ \  It follows from (\ref{e1}) and
(\ref{e2}) that
\begin{equation}
\ddot{q_{1}}=m_{2}\frac{q_{2}-q_{1}}{|
q_{2}-q_{1}|^{3}}+m_{3}\frac{q_{3}-q_{1}}{| q_{3}-q_{1}|^{3}}.
\end{equation}
Then by (\ref{e3})-(\ref{e5}), we obtain
\begin{equation}
\begin{aligned}
-\frac{4\pi^{2}}{T^{2}}q_{1}&=\frac{1}{l^{3}}(m_{2}q_{2}+m_{3}q_{3}-m_{2}q_{1}-m_{3}q_{1})\\
&=\frac{1}{l^{3}}(-m_{1}q_{1}-m_{2}q_{1}-m_{3}q_{1}),
\end{aligned}
\end{equation}
which implies
\begin{equation}
l^{3}=\frac{MT^{2}}{4\pi^{2}},\ \ \ \ \ \
\end{equation}
that is,
\begin{equation}\label{e17}
l=\sqrt[3]{\frac{MT^{2}}{4\pi^{2}}}.\ \ \ \Box
\end{equation}

\vspace{0.4cm}{\textbf{Lemma 2.6}}\ \ The radius $r_{1},r_{2},r_{3}$
of the planar circular orbits for the masses $m_{1},m_{2},m_{3}$ are
\begin{equation*}
r_{1}=\frac{\sqrt{m^{2}_{2}+m_{2}m_{3}+m^{2}_{3}}}{M}l,\ \ \ \ \ \
\end{equation*}
\begin{equation*}
r_{2}=\frac{\sqrt{m^{2}_{1}+m_{1}m_{3}+m^{2}_{3}}}{M}l,\ \ \ \ \ \
\end{equation*}
\begin{equation*}
r_{3}=\frac{\sqrt{m^{2}_{1}+m_{1}m_{2}+m^{2}_{2}}}{M}l.\ \ \ \ \ \
\end{equation*}

\vspace{0.4cm}{\textbf{Proof.}}\ \  Choose the geometrical center of
the initial configuration ($q_{1}(0),q_{2}(0),q_{3}(0)$) as the
origin of the coordinate (x,y). Without loss of generality, by
(\ref{e5}), we suppose the location coordinates of
$q_{1}(0),q_{2}(0),q_{3}(0)$ are
$A_{1}(\frac{\sqrt{3}l}{3},0),A_{2}(-\frac{\sqrt{3}l}{6},\frac{l}{2}),
A_{3}(-\frac{\sqrt{3}l}{6},-\frac{l}{2})$. Then we can get the
coordinate of the center of masses $m_{1},m_{2},m_{3}$ is
$C(\frac{\frac{\sqrt{3}}{3}m_{1}l-\frac{\sqrt{3}}{6}m_{2}l
-\frac{\sqrt{3}}{6}m_{3}l}{M},\frac{\frac{m_{2}}{2}l-\frac{m_{3}}{2}l}{M})$.
To make sure the Assumption (\ref{e4}) holds, we introduce the new
coordinate
\begin{equation*}
\left\{\begin{array}{ll} X=x-
\frac{\frac{\sqrt{3}}{3}m_{1}l-\frac{\sqrt{3}}{6}m_{2}l
-\frac{\sqrt{3}}{6}m_{3}l}{M},\\
Y=y-\frac{\frac{m_{2}}{2}l-\frac{m_{3}}{2}l}{M} .
\end{array}\right.
\end{equation*}
Hence in the new coordinate (X,Y), the location coordinates of
$q_{1}(0),q_{2}(0),q_{3}(0)$ are
$A_{1}(\frac{\frac{\sqrt{3}}{2}m_{2}l+\frac{\sqrt{3}}{2}m_{3}l}{M},$
$\frac{-\frac{m_{2}}{2}l+\frac{m_{3}}{2}l}{M}),$
$A_{2}(-\frac{\frac{\sqrt{3}}{2}m_{1}l}{M},\frac{\frac{m_{1}}{2}l+m_{3}l}{M}),
A_{3}(-\frac{\frac{\sqrt{3}}{2}m_{1}l}{M},-\frac{\frac{m_{1}}{2}l+m_{2}l}{M})$
and the center of masses $m_{1},m_{2},m_{3}$ is at the origin
$O(0,0)$. Then compared with (\ref{e3}), we have
\begin{equation}
r_{1}=|A_{1}O|=\frac{\sqrt{m^{2}_{2}+m_{2}m_{3}+m^{2}_{3}}}{M}l,\ \
\ \ \ \
\end{equation}
\begin{equation}
r_{2}=|A_{2}O|=\frac{\sqrt{m^{2}_{1}+m_{1}m_{3}+m^{2}_{3}}}{M}l,\ \
\ \ \ \
\end{equation}
\begin{equation}
r_{3}=|A_{3}O|=\frac{\sqrt{m^{2}_{1}+m_{1}m_{2}+m^{2}_{2}}}{M}l,\ \
\ \ \ \
\end{equation}
and
\begin{equation}
\sin\theta_{1}=\frac{-m_{2}+m_{3}}{2\sqrt{m^{2}_{2}+m_{2}m_{3}+m^{2}_{3}}},\
\ \ \ \ \ \
\cos\theta_{1}=\frac{\sqrt{3}(m_{2}+m_{3})}{2\sqrt{m^{2}_{2}+m_{2}m_{3}+m^{2}_{3}}},\
\ \ \ \ \ \
\end{equation}
\begin{equation}
\sin\theta_{2}=\frac{m_{1}+2m_{3}}{2\sqrt{m^{2}_{1}+m_{1}m_{3}+m^{2}_{3}}},\
\ \ \ \ \ \
\cos\theta_{2}=-\frac{\sqrt{3}m_{1}}{2\sqrt{m^{2}_{1}+m_{1}m_{3}+m^{2}_{3}}},\
\ \ \ \
\end{equation}
\begin{equation}
\sin\theta_{3}=-\frac{m_{1}+2m_{2}}{2\sqrt{m^{2}_{1}+m_{1}m_{2}+m^{2}_{2}}},\
\ \ \ \
\cos\theta_{3}=-\frac{\sqrt{3}m_{1}}{2\sqrt{m^{2}_{1}+m_{1}m_{2}+m^{2}_{2}}}.\
\ \ \Box
\end{equation}

\section{Proof of Theorems}
\ \ \ \ \ \ In order to get Theorems, we need two steps to complete
the proof.

Step 1: We will establish the existence of variational minimizers of
$f(q)$ in (\ref{e8}) on $\bar{\Lambda}_{\pm}$.

\vspace{0.4cm}{\textbf{Lemma 3.1}}\ \ $f(q)$ in (\ref{e8}) attains
its infimum on $\bar{\Lambda}_{\pm}$.

\vspace{0.4cm}{\textbf{Proof.}}\ \ By using Lemma 2.2, for $\forall
q\in \Lambda_{\pm}$, we can get that the equivalent norm of
(\ref{e7}) in $\bar{\Lambda}_{\pm}$ is
\begin{equation}
\|q\|\cong\Big[\int_{0}^{T}|\dot{q}|^{2}dt\Big]^{\frac{1}{2}}.
\end{equation}
Hence by the definition of $f(q)$, $f$ is coercive on
$\bar{\Lambda}_{\pm}$. Next, we claim that $f$ is weakly lower
semi-continuous on $\bar{\Lambda}_{\pm}$. In fact, for $\forall
q^{k}\in \Lambda_{\pm}$, if $q^{k}\rightharpoonup q$ weakly, by
compact embedding theorem, we have the uniformly convergence:
\begin{equation}
\max\limits_{0\leq t\leq T}|q^{k}(t)-q(t)|\rightarrow 0,\ \ \ \
k\rightarrow\infty,
\end{equation}
which implies
\begin{equation}\label{e9}
\int_{0}^{T}\sum\limits_{i=1}^{3}\frac{m_{i}}{|
q^{k}-q_{i}|}dt\rightarrow\int_{0}^{T}\sum\limits_{i=1}^{3}\frac{m_{i}
}{| q-q_{i}|}dt.
\end{equation}
It is well-known that the norm and its square are weakly lower
semi-continuous. Therefore, combined with (\ref{e9}), we obtain
\begin{equation}
\liminf\limits_{k\rightarrow\infty}f(q^{k})\geq f(q),
\end{equation}
that is, $f$ is weakly lower semi-continuous on
$\bar{\Lambda}_{\pm}$. By Lemma 2.1, we can get that $f(q)$ in
(\ref{e8}) attains its infimum on $\bar{\Lambda}_{\pm}$.\ \ $\Box$

Step 2: We will prove the variational minimizers in Lemma 3.1 is the
noncollision T-period solution of (\ref{e6}).

For any collision generalized solution $q$, we can estimate the
lower bound for the value of Lagrangian action functional.

\vspace{0.4cm}{\textbf{Lemma 3.2}}\ \ For
$\partial\Lambda_{\pm}=\{q\in
W^{1,2}(R/TZ,R^{2})|q(t+\frac{T}{2})=-q(t),\ \exists 1\leq
i^{\pm}_{0}\leq 3, t_{i^{\pm}_{0}}\in[0,T] \ s.t. \
q_{i^{\pm}_{0}}(t_{i^{\pm}_{0}})=q(t_{i^{\pm}_{0}})\}$, we have
\begin{equation*}
\inf\limits_{q\in\partial\Lambda_{\pm}}f(q)\geq\frac{3}{2}(2\pi)^{2/3}C
M^{-1/3}T^{1/3} \triangleq d_{1},
\end{equation*}
where
\begin{equation*}
C=\min\left\{
\begin{array}{c}
2^{\frac{2}{3}}m_{1}+m_{2}+m_{3}-\frac{1}{3M}(m_{1}m_{2}+m_{1}m_{3}+m_{2}m_{3}),\\
2^{\frac{2}{3}}m_{2}+m_{1}+m_{3}-\frac{1}{3M}(m_{1}m_{2}+m_{1}m_{3}+m_{2}m_{3}),\\
2^{\frac{2}{3}}m_{3}+m_{1}+m_{2}-\frac{1}{3M}(m_{1}m_{2}+m_{1}m_{3}+m_{2}m_{3})
\end{array}
\right\}.
\end{equation*}

\vspace{0.4cm}{\textbf{Proof.}}\ \ It follows from (\ref{e4}) that
\begin{equation}
\sum\limits_{i=1}^{3}m_{i}\dot{q}_{i}=0,
\end{equation}
which implies
\begin{eqnarray}
\sum\limits_{i=1}^{3}m_{i}|\dot{q}-\dot{q}_{i}|^{2}&=&\sum\limits_{i=1}^{3}m_{i}
\Big(|\dot{q}|^{2}+|\dot{q}_{i}|^{2}-2\langle\dot{q},\dot{q}_{i}\rangle\Big)\nonumber\\
&=&M|\dot{q}|^{2}+\sum\limits_{i=1}^{3}m_{i}|\dot{q}_{i}|^{2}-2\Big\langle\dot{q},
\sum\limits_{i=1}^{3}m_{i}\dot{q}_{i}\Big\rangle\nonumber\\
&=&M|\dot{q}|^{2}+\sum\limits_{i=1}^{3}m_{i}|\dot{q}_{i}|^{2}.
\end{eqnarray}
Therefore
\begin{equation}
|\dot{q}|^{2}=\frac{1}{M}\sum\limits_{i=1}^{3}m_{i}\Big(|\dot{q}-\dot{q}_{i}|^{2}-|\dot{q}_{i}|^{2}\Big).
\end{equation}
Hence
\begin{eqnarray}
f(q)&=&\int_{0}^{T}\Big[\frac{1}{2}|\dot{q}|^{2}+\sum\limits_{i=1}^{3}\frac{m_{i}
}{| q-q_{i}|}\Big]dt\nonumber\\
&=&\frac{1}{M}\int_{0}^{T}\sum\limits_{i=1}^{3}m_{i}\Big[\frac{1}{2}|\dot{q}-\dot{q}_{i}|^{2}
+\frac{M}{|q-q_{i}|}\Big]dt-\frac{1}{2M}\int_{0}^{T}\sum\limits_{i=1}^{3}m_{i}|\dot{q}_{i}|^{2}dt.
\end{eqnarray}
If $q\in\bar{\Lambda}_{-}$ is a collision generalized solution, then
there exists $t_{i^{-}_{0}}\in [0,T]$ and $1\leq i^{-}_{0}\leq 3$
such that $q(t_{i^{-}_{0}}) = q_{i^{-}_{0}}(t_{i^{-}_{0}})$. Since
$q_{i}(t+\frac{T}{2})=-q_{i}(t)$, we obtain
$q(t_{i^{-}_{0}}+\frac{kT}{2})=q_{i^{-}_{0}}(t_{i^{-}_{0}}+\frac{kT}{2}),
\ \ \forall 0\leq k\leq 2$. So, by (1) of Lemma 2.4, we get
\begin{eqnarray}\label{e10}
\frac{1}{M}\int_{0}^{T}m_{i^{-}_{0}}\Big[\frac{1}{2}|\dot{q}-\dot{q}_{i^{-}_{0}}|^{2}
+\frac{M}{|q-q_{i^{-}_{0}}|}\Big]dt&=&\frac{2}{M}m_{i^{-}_{0}}\int_{0}^{\frac{T}{2}}\Big[\frac{1}{2}|\dot{q}-\dot{q}_{i^{-}_{0}}|^{2}
+\frac{M}{|q-q_{i^{-}_{0}}|}\Big]dt\nonumber\\
&\geq&\frac{3}{2}(2\pi)^{2/3}2^{2/3}m_{i^{-}_{0}}M^{-1/3}T^{1/3}.
\end{eqnarray}
For noncollision pair $q,q_{i}(i\neq i^{-}_{0})$, we have
$\int_{0}^{T}q(t)dt=0$, $\int_{0}^{T}q_{i}(t)dt=0$. Therefore
$\int_{0}^{T}\big(q(t)-q_{i}(t)\big)dt=0$. Hence by (2) of Lemma
2.4, we can get
\begin{equation}\label{e11}
\frac{1}{M}\int_{0}^{T}\sum\limits_{i\neq
i^{-}_{0}}m_{i}\Big[\frac{1}{2}|\dot{q}-\dot{q}_{i}|^{2}
+\frac{M}{|q-q_{i}|}\Big]dt\geq\frac{3}{2}(2\pi)^{2/3}(M-m_{i^{-}_{0}})M^{-1/3}T^{1/3}.
\end{equation}
For the other term of $f$, using the expression for the orbits
$q_{1},q_{2},q_{3}$ as in (\ref{e3}), Lemma 2.5 and Lemma 2.6, we
obtain
\begin{equation}\label{e12}
-\frac{1}{2M}\int_{0}^{T}\sum\limits_{i=1}^{3}m_{i}|\dot{q}_{i}|^{2}dt
=-\frac{1}{2}(2\pi)^{2/3}(m_{1}m_{2}+m_{1}m_{3}+m_{2}m_{3})M^{-4/3}T^{1/3}.
\end{equation}
Therefore, it follows from (\ref{e10}) - (\ref{e12}) that
\begin{equation}
\inf\limits_{q\in\partial\Lambda_{-}}f(q)\geq\frac{3}{2}(2\pi)^{2/3}C
M^{-1/3}T^{1/3} \triangleq d_{1},
\end{equation}
where
\begin{equation*}
C=\min\left\{
\begin{array}{c}
2^{\frac{2}{3}}m_{1}+m_{2}+m_{3}-\frac{1}{3M}(m_{1}m_{2}+m_{1}m_{3}+m_{2}m_{3}),\\
2^{\frac{2}{3}}m_{2}+m_{1}+m_{3}-\frac{1}{3M}(m_{1}m_{2}+m_{1}m_{3}+m_{2}m_{3}),\\
2^{\frac{2}{3}}m_{3}+m_{1}+m_{2}-\frac{1}{3M}(m_{1}m_{2}+m_{1}m_{3}+m_{2}m_{3})
\end{array}
\right\}.
\end{equation*}
Similarly, if $q\in\bar{\Lambda}_{+}$ is a collision generalized
solution, we have
\begin{equation}
\inf\limits_{q\in\partial\Lambda_{+}}f(q)\geq\frac{3}{2}(2\pi)^{2/3}C
M^{-1/3}T^{1/3} \triangleq d_{1},\ \ \Box
\end{equation}

\vspace{0.4cm}\textbf{Proof of Theorem 1.1}\ \ In order to get
Theorem 1.1, we are going to find a test loop
$\tilde{q}\in\Lambda_{-}$ such that $f(\tilde{q})\leq d_{2}$. Then
the minimizer of $f$ on $\bar{\Lambda}_{-}$ must be a noncollision
solution if $d_{2}<d_{1}$.

\

Let $a > 0, b>0$, $\theta\in[0,2\pi)$ and
\begin{equation}
\tilde{q}-q_{1}=\bigg(a\cos\Big(-\frac{2\pi}{T}
t+\theta\Big),b\sin\Big(-\frac{2\pi}{T} t+\theta\Big)\bigg)^{T}.\ \
\ \ \ \ \ \ \ \ \ \ \ \ \ \ \ \ \ \ \ \ \ \ \ \ \ \ \ \ \ \ \ \ \ \
\ \ \ \ \ \ \ \ \ \
\end{equation}
Hence
\begin{eqnarray}
\tilde{q}-q_{2}&=&\tilde{q}-q_{1}+q_{1}-q_{2}\nonumber\\
&=&(q_{1}-q_{2})+(\tilde{q}-q_{1})\nonumber\\
&=&\bigg(r_{1}\cos\Big(\frac{2\pi}{T}
t+\theta_{1}\Big)-r_{2}\cos\Big(\frac{2\pi}{T}
t+\theta_{2}\Big)+a\cos\Big(-\frac{2\pi}{T}
t+\theta\Big),r_{1}\sin\Big(\frac{2\pi}{T}
t+\theta_{1}\Big)\ \ \ \ \ \ \ \nonumber\\
& &-r_{2}\sin\Big(\frac{2\pi}{T}
t+\theta_{2}\Big)+b\sin\Big(-\frac{2\pi}{T} t+\theta\Big)\bigg)^{T},
\end{eqnarray}
\begin{eqnarray}
\tilde{q}-q_{3}&=&\bigg(r_{1}\cos\Big(\frac{2\pi}{T}
t+\theta_{1}\Big)-r_{3}\cos\Big(\frac{2\pi}{T}
t+\theta_{3}\Big)+a\cos\Big(-\frac{2\pi}{T}
t+\theta\Big),r_{1}\sin\Big(\frac{2\pi}{T}
t+\theta_{1}\Big)\ \ \ \ \ \ \ \nonumber\\
& &-r_{3}\sin\Big(\frac{2\pi}{T}
t+\theta_{3}\Big)+b\sin\Big(-\frac{2\pi}{T} t+\theta\Big)\bigg)^{T}.
\end{eqnarray}

\

It is easy to see that $\tilde{q}\in\Lambda_{-}$ and
\begin{eqnarray}\label{e13}
|\dot{\tilde{q}}-\dot{q_{1}}|^{2}&=&\Big(\frac{2\pi}{T}\Big)^{2}\bigg[\frac{a^{2}+b^{2}}{2}-\frac{a^{2}-b^{2}}{2}cos\Big(\frac{4\pi}{T}
t-2\theta\Big)\bigg],\ \ \ \ \ \ \ \ \ \ \ \ \ \ \ \ \ \ \ \ \ \ \ \
\ \ \ \ \ \ \ \ \ \ \ \ \
\end{eqnarray}
\begin{eqnarray}\label{14}
|\tilde{q}-q_{1}|=\sqrt{\frac{a^{2}+b^{2}}{2}+\frac{a^{2}-b^{2}}{2}cos\Big(\frac{4\pi}{T}
t-2\theta\Big)},\ \ \ \ \ \ \ \ \ \ \ \ \ \ \ \ \ \ \ \ \ \ \ \ \ \
\ \ \ \ \ \ \ \ \ \ \ \ \
\end{eqnarray}
\begin{eqnarray}\label{e15}
|\dot{\tilde{q}}-\dot{q_{2}}|^{2}&=&\Big(\frac{2\pi}{T}\Big)^{2}\bigg\{\frac{a^{2}+b^{2}}{2}-\frac{a^{2}-b^{2}}{2}cos\Big(\frac{4\pi}{T}
t-2\theta\Big)+r_{1}^{2}+r_{2}^{2}-2r_{1}r_{2}cos(\theta_{2}-\theta_{1})\nonumber\\
& &-(a+b)\Big[r_{1}cos\Big(\frac{4\pi}{T} t+\theta_{1}-\theta\Big)
-r_{2}cos\Big(\frac{4\pi}{T}
t+\theta_{2}-\theta\Big)\Big]\nonumber\\
&
&+(a-b)\big[r_{1}cos(\theta_{1}+\theta)-r_{2}cos(\theta_{2}+\theta)\big]\bigg\},
\end{eqnarray}
\begin{eqnarray}\label{e16}
|\tilde{q}-q_{2}|&=&\bigg\{\frac{a^{2}+b^{2}}{2}+\frac{a^{2}-b^{2}}{2}cos\Big(\frac{4\pi}{T}
t-2\theta\Big)+r_{1}^{2}+r_{2}^{2}-2r_{1}r_{2}cos(\theta_{2}-\theta_{1})\ \ \ \ \ \ \ \ \nonumber\\
& &+(a+b)\Big[r_{1}cos\Big(\frac{4\pi}{T} t+\theta_{1}-\theta\Big)
-r_{2}cos\Big(\frac{4\pi}{T}
t+\theta_{2}-\theta\Big)\Big]\nonumber\\
&
&+(a-b)\big[r_{1}cos(\theta_{1}+\theta)-r_{2}cos(\theta_{2}+\theta)\big]\bigg\}^{\frac{1}{2}},
\end{eqnarray}
\begin{eqnarray}\label{e17}
|\dot{\tilde{q}}-\dot{q_{3}}|^{2}&=&\Big(\frac{2\pi}{T}\Big)^{2}\bigg\{\frac{a^{2}+b^{2}}{2}-\frac{a^{2}-b^{2}}{2}cos\Big(\frac{4\pi}{T}
t-2\theta\Big)+r_{1}^{2}+r_{3}^{2}-2r_{1}r_{3}cos(\theta_{3}-\theta_{1})\nonumber\\
& &-(a+b)\Big[r_{1}cos\Big(\frac{4\pi}{T} t+\theta_{1}-\theta\Big)
-r_{3}cos\Big(\frac{4\pi}{T}
t+\theta_{3}-\theta\Big)\Big]\nonumber\\
&
&+(a-b)\big[r_{1}cos(\theta_{1}+\theta)-r_{3}cos(\theta_{3}+\theta)\big]\bigg\},
\end{eqnarray}
\begin{eqnarray}\label{e18}
|\tilde{q}-q_{3}|&=&\bigg\{\frac{a^{2}+b^{2}}{2}+\frac{a^{2}-b^{2}}{2}cos\Big(\frac{4\pi}{T}
t-2\theta\Big)+r_{1}^{2}+r_{3}^{2}-2r_{1}r_{3}cos(\theta_{2}-\theta_{1})\ \ \ \ \ \ \ \ \nonumber\\
& &+(a+b)\Big[r_{1}cos\Big(\frac{4\pi}{T} t+\theta_{1}-\theta\Big)
-r_{3}cos\Big(\frac{4\pi}{T}
t+\theta_{3}-\theta\Big)\Big]\nonumber\\
&
&+(a-b)\big[r_{1}cos(\theta_{1}+\theta)-r_{3}cos(\theta_{3}+\theta)\big]\bigg\}^{\frac{1}{2}},
\end{eqnarray}
\begin{equation}\label{e19}
|\dot{q_{1}}|^{2}=\Big(\frac{2\pi}{T}\Big)^{2}r_{1}^{2},\ \ \ \
|\dot{q_{2}}|^{2}=\Big(\frac{2\pi}{T}\Big)^{2}r_{2}^{2},\ \ \ \
|\dot{q_{3}}|^{2}=\Big(\frac{2\pi}{T}\Big)^{2}r_{3}^{2}.
\end{equation}
Therefore by (\ref{e13})-(\ref{e19}), we get
\begin{eqnarray}\label{e20}
f(\tilde{q})&=&\frac{1}{M}\int_{0}^{T}\sum\limits_{i=1}^{3}m_{i}\Big[\frac{1}{2}|\dot{\tilde{q}}-\dot{q}_{i}|^{2}
+\frac{M}{|\tilde{q}-q_{i}|}\Big]dt-\frac{1}{2M}\int_{0}^{T}\sum\limits_{i=1}^{3}m_{i}|\dot{q}_{i}|^{2}dt\nonumber\\
&=&\frac{2\pi^{2}}{T}\Big\{\frac{a^{2}+b^{2}}{2}+\frac{m_{2}+m_{3}-m_{1}}{M}r_{1}^{2}-\frac{2m_{2}r_{2}
cos(\theta_{2}-\theta_{1})+2m_{3}r_{3}cos(\theta_{3}-\theta_{1})}{M}r_{1}\nonumber\\
& &+\frac{m_{2}(a-b)}{M}
[r_{1}cos(\theta_{1}+\theta)-r_{2}cos(\theta_{2}+\theta)]+\frac{m_{3}(a-b)}{M}
[r_{1}cos(\theta_{1}+\theta)\nonumber\\
&
&-r_{3}cos(\theta_{3}+\theta)]\Big\}+m_{1}\int_{0}^{T}\bigg[\frac{a^{2}+b^{2}}{2}+\frac{a^{2}-b^{2}}{2}cos\Big(\frac{4\pi}{T}
t-2\theta\Big)\bigg]^{-\frac{1}{2}}dt\nonumber\\
& &+\sum\limits_{i=2}^{3}\int_{0}^{T}m_{i}
\bigg\{\frac{a^{2}+b^{2}}{2}+\frac{a^{2}-b^{2}}{2}cos\Big(\frac{4\pi}{T}
t-2\theta\Big)+r_{1}^{2}+r_{i}^{2}-2r_{1}r_{i}cos(\theta_{i}-\theta_{1})\nonumber\\
& &+(a+b)\Big[r_{1}cos\Big(\frac{4\pi}{T} t+\theta_{1}-\theta\Big)
-r_{i}cos\Big(\frac{4\pi}{T}
t+\theta_{i}-\theta\Big)\Big]\nonumber\\
&
&+(a-b)\big[r_{1}cos(\theta_{1}+\theta)-r_{i}cos(\theta_{i}+\theta)\big]\bigg\}^{-\frac{1}{2}}dt\nonumber\\
&=&d_{2}(a,b,\theta).
\end{eqnarray}

\

In order to estimate $d_{2}$, we have computed the numerical values
of $d_{2} = f(q)$ over some selected test loops. The computation of
the integral that appears in (\ref{e20}) has been done using the
function $\{quad\}$ of Mathematica 7.1 with an error less than
$10^{-6}$. Let $T=1$, the results of the numerical explorations are
given in Table 1 with $M=1$ and Table 2 with $m_{1}=m_{2}=m_{3}=1$.

\begin{table}[!hbt]
   \centering
   \caption{Parameters for test loops for Theoerm 1.1}
      \[
        \begin{tabular}{|c|c|c|c|c|c|c|c|c|}
        \hline
         a     & b   & $\theta$           &m1    &m2    &m3   & d1        &d2 \\ \hline
        0.13  &0.49  &$\frac{\pi}{20}$   &0.29  &0.42  &0.29 &5.419669   &5.417862 \\ \hline
        0.15  &0.49  &$\frac{\pi}{20}$   &0.29  &0.41  &0.30 &5.417626   &5.416591 \\ \hline
        0.15  &0.49  &$\frac{\pi}{20}$   &0.29  &0.42  &0.29 &5.419669   &5.413794 \\ \hline
        0.15  &0.49  &$\frac{\pi}{20}$   &0.30  &0.35  &0.35 &5.441499   &5.436767 \\ \hline
        0.15  &0.51  &$\frac{\pi}{20}$   &0.30  &0.36  &0.34 &5.441669   &5.437985  \\ \hline
        0.15  &0.51  &$\frac{\pi}{20}$   &0.30  &0.37  &0.33 &5.442180   &5.433615 \\ \hline
        0.15  &0.51  &$\frac{\pi}{20}$   &0.30  &0.38  &0.32 &5.443031   &5.429587 \\ \hline
        0.15  &0.51  &$\frac{\pi}{20}$   &0.30  &0.39  &0.31 &5.444223   &5.425898 \\ \hline
        0.15  &0.51  &$\frac{\pi}{20}$   &0.30  &0.40  &0.30 &5.445755   &5.422550 \\ \hline
        0.15  &0.53  &$\frac{\pi}{20}$   &0.31  &0.35  &0.34 &5.470820   &5.467576  \\ \hline
        0.15  &0.53  &$\frac{\pi}{20}$   &0.31  &0.36  &0.33 &5.471160   &5.462971 \\ \hline
        0.15  &0.53  &$\frac{\pi}{20}$   &0.31  &0.37  &0.32 &5.471841   &5.458707 \\ \hline
        0.15  &0.53  &$\frac{\pi}{20}$   &0.31  &0.38  &0.31 &5.472863   &5.454784 \\ \hline
        0.17  &0.45  &$\frac{\pi}{20}$   &0.32  &0.32  &0.36 &5.500992   &5.488608  \\ \hline
        0.17  &0.47  &$\frac{\pi}{20}$   &0.32  &0.33  &0.35 &5.500481   &5.454518  \\ \hline
        0.17  &0.47  &$\frac{\pi}{20}$   &0.32  &0.34  &0.34 &5.500311   &5.449987  \\ \hline
        0.17  &0.47  &$\frac{\pi}{20}$   &0.33  &0.34  &0.33 &5.530142   &5.444254  \\ \hline
        0.45  &0.15  &$\pi$              &0.33  &0.31  &0.36 &5.471160   &5.456006  \\ \hline
        0.45  &0.15  &$\pi$              &0.33  &0.32  &0.35 &5.500481   &5.455325  \\ \hline
        0.45  &0.15  &$\pi$              &0.33  &0.33  &0.34 &5.530142   &5.454984  \\ \hline
        0.47  &0.13  &$\pi$              &0.34  &0.30  &0.36 &5.441669   &5.439671  \\ \hline
        0.47  &0.13  &$\pi$              &0.34  &0.31  &0.35 &5.470820   &5.438820  \\ \hline
        0.47  &0.13  &$\pi$              &0.34  &0.32  &0.34 &5.500311   &5.438309  \\ \hline
        0.47  &0.15  &$\pi$              &0.35  &0.30  &0.35 &5.441499   &5.417900  \\ \hline
        0.49  &0.15  &$\pi$              &0.36  &0.29  &0.35 &5.412519   &5.411552  \\ \hline
        0.49  &0.15  &$\pi$              &0.36  &0.32  &0.32 &5.500992   &5.410020  \\ \hline
        0.49  &0.15  &$\pi$              &0.37  &0.29  &0.34 &5.412859   &5.411962  \\ \hline
        0.49  &0.15  &$\pi$              &0.37  &0.30  &0.33 &5.442180   &5.411281  \\ \hline
        0.49  &0.15  &$\pi$              &0.37  &0.31  &0.32 &5.471841   &5.410940  \\ \hline
        0.49  &0.15  &$\pi$              &0.38  &0.29  &0.33 &5.413540   &5.412712  \\ \hline
        0.49  &0.15  &$\pi$              &0.38  &0.30  &0.32 &5.443031   &5.412201  \\ \hline
        0.49  &0.15  &$\pi$              &0.38  &0.31  &0.31 &5.472863   &5.412031 \\ \hline
        0.49  &0.15  &$\pi$              &0.39  &0.29  &0.32 &5.414562   &5.413803  \\ \hline
        0.49  &0.15  &$\pi$              &0.39  &0.30  &0.31 &5.444223   &5.413462  \\ \hline
        0.49  &0.17  &$\pi$              &0.40  &0.29  &0.31 &5.415924   &5.415807  \\ \hline
        0.49  &0.17  &$\pi$              &0.40  &0.30  &0.30 &5.445755   &5.415637  \\ \hline
        0.49  &0.17  &$\pi$              &0.41  &0.30  &0.29 &5.417626   &5.416078 \\ \hline
        0.49  &0.17  &$\pi$              &0.42  &0.29  &0.29 &5.419669   &5.416689 \\ \hline

   \end {tabular}
   \]
\label{Table 1}
\end{table}

\begin{table}[!hbt]
   \centering
   \caption{Parameters for test loops for Theoerm 1.1}
      \[
        \begin{tabular}{|c|c|c|c|c|c|c|c|c|}
        \hline
         a     & b   & $\theta$           &m1    &m2    &m3   & d1        &d2 \\ \hline
        0.15  &0.67  &$\frac{\pi}{30}$   &1.00  &1.00  &1.00 &11.523843   &11.505860 \\ \hline
        0.15  &0.67  &$\frac{\pi}{30}$   &1.00  &1.00  &1.00 &11.523843   &11.505860 \\ \hline
        0.15  &0.69  &$\frac{\pi}{30}$   &1.00  &1.00  &1.00 &11.523843   &11.444212 \\ \hline
        0.17  &0.65  &$\frac{\pi}{30}$   &1.00  &1.00  &1.00 &11.523843   &11.493238 \\ \hline
        0.17  &0.67  &$\frac{\pi}{20}$   &1.00  &1.00  &1.00 &11.523843   &11.452135 \\ \hline
        0.17  &0.69  &$\frac{\pi}{20}$   &1.00  &1.00  &1.00 &11.523843   &11.400124 \\ \hline
        0.19  &0.63  &$\frac{\pi}{30}$   &1.00  &1.00  &1.00 &11.523843   &11.519350 \\ \hline
        0.19  &0.65  &$\frac{\pi}{20}$   &1.00  &1.00  &1.00 &11.523843   &11.455969 \\ \hline
        0.19  &0.67  &$\frac{\pi}{20}$   &1.00  &1.00  &1.00 &11.523843   &11.386608 \\ \hline
        0.19  &0.69  &$\frac{\pi}{20}$   &1.00  &1.00  &1.00 &11.523843   &11.344747 \\ \hline
        0.61  &0.23  &$\pi$              &1.00  &1.00  &1.00 &11.523843   &11.516685 \\ \hline
        0.63  &0.19  &$\pi$              &1.00  &1.00  &1.00 &11.523843   &11.489791 \\ \hline
        0.63  &0.21  &$\pi$              &1.00 &1.00  &1.00 &11.523843   &11.436105 \\ \hline
        0.65  &0.17  &$\pi$              &1.00 &1.00  &1.00 &11.523843   &11.461786 \\ \hline
        0.65  &0.19  &$\pi$              &1.00 &1.00  &1.00 &11.523843   &11.392115 \\ \hline
        0.65  &0.21  &$\pi$              &1.00 &1.00  &1.00 &11.523843   &11.349366 \\ \hline
        0.67  &0.15  &$\pi$              &1.00 &1.00  &1.00 &11.523843   &11.472422 \\ \hline
        0.67  &0.17  &$\pi$              &1.00 &1.00  &1.00 &11.523843   &11.383978 \\ \hline
        0.67  &0.19  &$\pi$              &1.00 &1.00  &1.00 &11.523843   &11.324970 \\ \hline
        0.67  &0.21  &$\pi$              &1.00 &1.00  &1.00 &11.523843   &11.291915 \\ \hline
        0.69  &0.13  &$\pi$              &1.00 &1.00  &1.00 &11.523843   &11.522980 \\ \hline
        0.69  &0.15  &$\pi$              &1.00 &1.00  &1.00 &11.523843   &11.412094 \\ \hline
        0.69  &0.17  &$\pi$              &1.00 &1.00  &1.00 &11.523843   &11.334189 \\ \hline
        0.69  &0.19  &$\pi$              &1.00 &1.00  &1.00 &11.523843   &11.284714 \\ \hline

   \end {tabular}
   \]
\label{Table 2}
\end{table}

\newpage

For the parameters $a,b,\theta$ given in Table 1 and Table 2, we all
have $d_{2}< d_{1}$. This completes the Proof of Theorem 1.1.\ \
$\Box$

\vspace{0.4cm}\textbf{Proof of Theorem 1.2}\ \ To get Theorem 1.2,
we are going to find a test loop $\bar{q}\in\Lambda_{+}$ such that
$f(\bar{q})\leq d_{3}$. Then the minimizer of $f$ on
$\bar{\Lambda}_{+}$ must be a noncollision solution if
$d_{3}<d_{1}$.

\

Let $a > 0$, $\theta\in[0,2\pi)$ and
\begin{equation}
\bar{q}-q_{1}=ae^{\sqrt{-1}(\frac{2\pi}{T} t+\theta)}.\ \ \ \ \ \ \
\ \ \ \ \ \ \ \ \ \ \ \ \ \ \ \ \ \ \ \ \ \ \ \ \ \ \ \ \ \ \ \ \ \
\ \ \ \
\end{equation}
Hence
\begin{eqnarray}
\bar{q}-q_{2}&=&q_{1}+ae^{\sqrt{-1}(\frac{2\pi}{T} t+\theta)}-q_{2}\nonumber\\
&=&r_{1}e^{\sqrt{-1}(\frac{2\pi}{T}
t+\theta_{1})}-r_{2}e^{\sqrt{-1}(\frac{2\pi}{T}
t+\theta_{2})}+ae^{\sqrt{-1}(\frac{2\pi}{T} t+\theta)},
\end{eqnarray}
\begin{eqnarray}
\bar{q}-q_{3}&=&q_{1}+ae^{\sqrt{-1}(\frac{2\pi}{T} t+\theta)}-q_{3}\nonumber\\
&=&r_{1}e^{\sqrt{-1}(\frac{2\pi}{T}
t+\theta_{1})}-r_{3}e^{\sqrt{-1}(\frac{2\pi}{T}
t+\theta_{3})}+ae^{\sqrt{-1}(\frac{2\pi}{T} t+\theta)}.
\end{eqnarray}

\

It is easy to see that $\bar{q}\in\Lambda_{+}$ and
\begin{eqnarray}\label{e21}
|\dot{\bar{q}}-\dot{q_{1}}|^{2}&=&\Big(\frac{2\pi}{T}\Big)^{2}a^{2},\
\ \ \ |\bar{q}-q_{1}|=a,\ \ \ \ \ \ \ \ \ \ \ \ \ \ \ \ \ \ \ \ \ \
\end{eqnarray}

\begin{eqnarray}\label{e22}
|\dot{\bar{q}}-\dot{q_{2}}|^{2}=\Big(\frac{2\pi}{T}\Big)^{2}\big[a^{2}+r_{1}^{2}+
r_{2}^{2}-2r_{1}r_{2}cos(\theta_{2}-\theta_{1})+2ar_{1}cos(\theta_{1}-\theta)
-2ar_{2}cos(\theta_{2}-\theta)\big],
\end{eqnarray}
\begin{eqnarray}\label{e23}
|\bar{q}-q_{2}|=\big[a^{2}+r_{1}^{2}+
r_{2}^{2}-2r_{1}r_{2}cos(\theta_{2}-\theta_{1})+2ar_{1}cos(\theta_{1}-\theta)
-2ar_{2}cos(\theta_{2}-\theta)\big]^{\frac{1}{2}},\ \ \ \ \ \
\end{eqnarray}
\begin{eqnarray}\label{e24}
|\dot{\bar{q}}-\dot{q_{3}}|^{2}=\Big(\frac{2\pi}{T}\Big)^{2}\big[a^{2}+r_{1}^{2}+
r_{3}^{2}-2r_{1}r_{3}cos(\theta_{3}-\theta_{1})+2ar_{1}cos(\theta_{1}-\theta)
-2ar_{3}cos(\theta_{3}-\theta)\big],
\end{eqnarray}
\begin{eqnarray}\label{e25}
|\bar{q}-q_{3}|=\big[a^{2}+r_{1}^{2}+
r_{3}^{2}-2r_{1}r_{3}cos(\theta_{3}-\theta_{1})+2ar_{1}cos(\theta_{1}-\theta)
-2ar_{3}cos(\theta_{3}-\theta)\big]^{\frac{1}{2}},\ \ \ \ \ \
\end{eqnarray}
\begin{equation}\label{e26}
|\dot{q_{1}}|^{2}=\Big(\frac{2\pi}{T}\Big)^{2}r_{1}^{2},\ \ \ \
|\dot{q_{2}}|^{2}=\Big(\frac{2\pi}{T}\Big)^{2}r_{2}^{2},\ \ \ \
|\dot{q_{3}}|^{2}=\Big(\frac{2\pi}{T}\Big)^{2}r_{3}^{2}.
\end{equation}

Therefore by (\ref{e21})-(\ref{e26}), we get
\begin{eqnarray}\label{e27}
f(\bar{q})&=&\frac{1}{M}\int_{0}^{T}\sum\limits_{i=1}^{3}m_{i}\Big[\frac{1}{2}|\dot{\bar{q}}-\dot{q}_{i}|^{2}
+\frac{M}{|\bar{q}-q_{i}|}\Big]dt-\frac{1}{2M}\int_{0}^{T}\sum\limits_{i=1}^{3}m_{i}|\dot{q}_{i}|^{2}dt\nonumber\\
&=&\frac{2\pi^{2}}{T}\Big[a^{2}+\frac{m_{2}+m_{3}-m_{1}}{M}r_{1}^{2}-\frac{2m_{2}r_{2}
cos(\theta_{2}-\theta_{1})+2m_{3}r_{3}cos(\theta_{3}-\theta_{1})}{M}r_{1}\nonumber\\
&
&+\frac{2(m_{2}+m_{3})}{M}ar_{1}cos(\theta_{1}-\theta)-\frac{2m_{2}r_{2}
cos(\theta_{2}-\theta)+2m_{3}r_{3}cos(\theta_{3}-\theta)}{M}a\Big]\nonumber\\
& &+\frac{m_{1}T}{a}+m_{2}\int_{0}^{T} \big[a^{2}+r_{1}^{2}+
r_{2}^{2}-2r_{1}r_{2}cos(\theta_{2}-\theta_{1})+2ar_{1}cos(\theta_{1}-\theta)\nonumber\\
&
&-2ar_{2}cos(\theta_{2}-\theta)\big]^{-\frac{1}{2}}dt+m_{3}\int_{0}^{T}
\big[a^{2}+r_{1}^{2}+
r_{3}^{2}-2r_{1}r_{3}cos(\theta_{3}-\theta_{1})\nonumber\\
& &+2ar_{1}cos(\theta_{1}-\theta)
-2ar_{3}cos(\theta_{3}-\theta)\big]^{-\frac{1}{2}}dt\nonumber\\
&=&d_{3}(a,\theta).
\end{eqnarray}

\

In order to estimate $d_{3}$, we have computed the numerical values
of $d_{3} = f(q)$ over some selected test loops. The computation of
the integral that appears in (\ref{e27}) has been done using the
function $\{quad\}$ of Mathematica 7.1 with an error less than
$10^{-6}$. Let $T=1$, the results of the numerical explorations are
given in Table 3 with $M=1$ and Table 4 with $m_{1}=m_{2}=m_{3}=1$.

\newpage

\begin{table}[!hbt]
   \centering
   \caption{Parameters for test loops for Theoerm 1.2}
      \[
        \begin{tabular}{|c|c|c|c|c|c|c|c|c|}
        \hline
         a       & $\theta$          &m1    &m2    &m3     & d1       &d3 \\ \hline
         0.17    &$\frac{\pi}{2}$    &0.10  &0.75  &0.15 &5.062791   &5.060773 \\ \hline
         0.17    &$\frac{\pi}{2}$    &0.10  &0.77  &0.13 &5.083903   &5.071551 \\ \hline
         0.17    &$\frac{\pi}{2}$    &0.10  &0.78  &0.12 &5.094969   &5.077450 \\ \hline
         0.17    &$\frac{\pi}{2}$    &0.10  &0.80  &0.10 &5.118123   &5.090270 \\ \hline
         0.17    &$\frac{\pi}{2}$    &0.15  &0.53  &0.32 &5.051742   &5.050040 \\ \hline
         0.17    &$\frac{\pi}{2}$    &0.15  &0.57  &0.28 &5.068768   &5.046398 \\ \hline
         0.17    &$\frac{\pi}{2}$    &0.15  &0.60  &0.25 &5.085112   &5.047242 \\ \hline
         0.17    &$\frac{\pi}{2}$    &0.15  &0.65  &0.20 &5.119162   &5.055458 \\ \hline
         0.17    &$\frac{\pi}{2}$    &0.15  &0.70  &0.15 &5.161725   &5.072186 \\ \hline
         0.17    &$\frac{\pi}{2}$    &0.15  &0.72  &0.13 &5.121130   &5.081261 \\ \hline
         0.17    &$\frac{\pi}{2}$    &0.20  &0.31  &0.49 &5.176554   &5.175168 \\ \hline
         0.17    &$\frac{\pi}{2}$    &0.20  &0.35  &0.45 &5.167020   &5.144967 \\ \hline
         0.17    &$\frac{\pi}{2}$    &0.20  &0.40  &0.40 &5.162763   &5.114876 \\ \hline
         0.17    &$\frac{\pi}{2}$    &0.20  &0.50  &0.30 &5.179789   &5.080232 \\ \hline
         0.17    &$\frac{\pi}{2}$    &0.20  &0.55  &0.25 &5.201070   &5.075680 \\ \hline
         0.17    &$\frac{\pi}{2}$    &0.20  &0.60  &0.20 &5.230864   &5.079639 \\ \hline
         0.19    &$\frac{\pi}{2}$    &0.25  &0.22  &0.53 &5.249837   &5.237465 \\ \hline
         0.19    &$\frac{\pi}{2}$    &0.25  &0.25  &0.50 &5.325541   &5.202291 \\ \hline
         0.19    &$\frac{\pi}{2}$    &0.25  &0.30  &0.45 &5.308516   &5.150479 \\ \hline
         0.19    &$\frac{\pi}{2}$    &0.25  &0.35  &0.40 &5.300003   &5.107178 \\ \hline
         0.19    &$\frac{\pi}{2}$    &0.25  &0.62  &0.13 &5.041112   &5.020454 \\ \hline
         0.19    &$\frac{\pi}{2}$    &0.30  &0.22  &0.48 &5.230258   &5.222385 \\ \hline
         0.19    &$\frac{\pi}{2}$    &0.30  &0.25  &0.45 &5.308516   &5.189765 \\ \hline
         0.19    &$\frac{\pi}{2}$    &0.30  &0.30  &0.40 &5.445755   &5.142208 \\ \hline
         0.19    &$\frac{\pi}{2}$    &0.30  &0.35  &0.35 &5.441499   &5.103164 \\ \hline
         0.19    &$\frac{\pi}{2}$    &0.30  &0.56  &0.14 &5.036553   &5.032137 \\ \hline
         0.21    &$\frac{\pi}{2}$    &0.35  &0.21  &0.44 &5.192935   &5.184596 \\ \hline
         0.21    &$\frac{\pi}{2}$    &0.35  &0.29  &0.36 &5.412519   &5.092092 \\ \hline
         0.21    &$\frac{\pi}{2}$    &0.35  &0.39  &0.26 &5.327621   &5.007107 \\ \hline
         0.21    &$\frac{\pi}{2}$    &0.35  &0.48  &0.17 &5.091316   &4.959734 \\ \hline
         0.21    &$\frac{\pi}{2}$    &0.35  &0.53  &0.12 &4.971952   &4.945333 \\ \hline
         0.21    &$\frac{\pi}{3}$    &0.40  &0.28  &0.32 &5.386433   &5.342981 \\ \hline
         0.21    &$\frac{\pi}{3}$    &0.40  &0.32  &0.28 &5.386433   &5.287294 \\ \hline
         0.21    &$\frac{\pi}{3}$    &0.40  &0.36  &0.24 &5.271874   &5.237055 \\ \hline
         0.21    &$\frac{\pi}{3}$    &0.40  &0.38  &0.22 &5.216638   &5.213978 \\ \hline
         0.23    &$\frac{\pi}{2}$    &0.45  &0.19  &0.36 &5.139742   &5.127834 \\ \hline
         0.23    &$\frac{\pi}{2}$    &0.45  &0.29  &0.26 &5.337836   &5.003006 \\ \hline
         0.23    &$\frac{\pi}{2}$    &0.45  &0.37  &0.18 &5.112805   &4.927660 \\ \hline
         0.23    &$\frac{\pi}{2}$    &0.45  &0.46  &0.09 &4.885693   &4.868944 \\ \hline
         0.23    &$\frac{\pi}{2}$    &0.50  &0.18  &0.32 &5.123871   &5.108878 \\ \hline
         0.23    &$\frac{\pi}{2}$    &0.50  &0.23  &0.27 &5.266218   &5.044762 \\ \hline
         0.23    &$\frac{\pi}{2}$    &0.50  &0.29  &0.21 &5.208258   &4.979058 \\ \hline
         0.23    &$\frac{\pi}{2}$    &0.50  &0.37  &0.13 &4.990036   &4.910522 \\ \hline
         0.23    &$\frac{\pi}{2}$    &0.50  &0.41  &0.09 &4.889098   &4.884426 \\ \hline

   \end {tabular}
   \]
\label{Table 3}
\end{table}

\begin{table}[!hbt]
   \centering
   \caption{Parameters for test loops for Theoerm 1.2}
      \[
        \begin{tabular}{|c|c|c|c|c|c|c|c|c|}
        \hline
         a       & $\theta$         &m1    &m2    &m3     & d1       &d3 \\ \hline
         0.21    &$\frac{\pi}{2}$   &1.00  &1.00  &1.00 &11.523843   &11.327950 \\ \hline
         0.23    &$\frac{\pi}{2}$   &1.00  &1.00  &1.00 &11.523843   &11.036769 \\ \hline
         0.23    &$\frac{\pi}{3}$   &1.00  &1.00  &1.00 &11.523843   &11.336568 \\ \hline
         0.25    &$\frac{\pi}{2}$   &1.00  &1.00  &1.00 &11.523843   &10.821272 \\ \hline
         0.25    &$\frac{\pi}{3}$   &1.00  &1.00  &1.00 &11.523843   &11.187475 \\ \hline
         0.25    &$\frac{\pi}{4}$   &1.00  &1.00  &1.00 &11.523843   &11.453195 \\ \hline
         0.27    &$\frac{\pi}{2}$   &1.00  &1.00  &1.00 &11.523843   &10.667031 \\ \hline
         0.27    &$\frac{\pi}{3}$   &1.00  &1.00  &1.00 &11.523843   &11.107374 \\ \hline
         0.27    &$\frac{\pi}{4}$   &1.00  &1.00  &1.00 &11.523843   &11.411685 \\ \hline
         0.29    &$\frac{\pi}{2}$   &1.00  &1.00  &1.00 &11.523843   &10.563849 \\ \hline
         0.29    &$\frac{\pi}{3}$   &1.00  &1.00  &1.00 &11.523843   &11.085761 \\ \hline
         0.29    &$\frac{\pi}{4}$   &1.00  &1.00  &1.00 &11.523843   &11.430090 \\ \hline
         0.31    &$\frac{\pi}{2}$   &1.00  &1.00  &1.00 &11.523843   &10.504424 \\ \hline
         0.31    &$\frac{\pi}{3}$   &1.00  &1.00  &1.00 &11.523843   &11.114860 \\ \hline
         0.31    &$\frac{\pi}{4}$   &1.00  &1.00  &1.00 &11.523843   &11.500414 \\ \hline
         0.33    &$\frac{\pi}{2}$   &1.00  &1.00  &1.00 &11.523843   &10.483477 \\ \hline
         0.33    &$\frac{\pi}{3}$   &1.00  &1.00  &1.00 &11.523843   &11.188786 \\ \hline
         0.35    &$\frac{\pi}{2}$   &1.00  &1.00  &1.00 &11.523843   &10.497161 \\ \hline
         0.35    &$\frac{\pi}{3}$   &1.00  &1.00  &1.00 &11.523843   &11.302997 \\ \hline
         0.37    &$\frac{\pi}{2}$   &1.00  &1.00  &1.00 &11.523843   &10.542652 \\ \hline
         0.37    &$\frac{\pi}{3}$   &1.00  &1.00  &1.00 &11.523843   &11.453926 \\ \hline
         0.39    &$\frac{\pi}{2}$   &1.00  &1.00  &1.00 &11.523843   &10.617860 \\ \hline

   \end {tabular}
   \]
\label{Table 4}
\end{table}

\

For the parameters $a,\theta$ given in Table 3 and Table 4, we all
have $d_{3}< d_{1}$. This completes the Proof of Theorem 1.2.\ \
$\Box$

\end{document}